\documentclass[%
 reprint,
 amsmath,amssymb,
 aps,
]{revtex4-2}

\usepackage{graphicx}
\usepackage{dcolumn}
\usepackage{bm}

\usepackage{tikz}
\usetikzlibrary{positioning}
\usetikzlibrary{calc,patterns,decorations.pathmorphing,decorations.markings}
\usepackage{stanli} 
\usetikzlibrary{calc,intersections,patterns}
\usepackage [ europeanresistors ]{ circuitikz }
\usepackage{physics}
\usepackage{url}

\begin{document}

\preprint{APS/123-QED}

\title{Paradox with Phase-Coupled Interferometers}

\author{Saba Etezad-Razavi}
\author{Lucien Hardy}%
 \email{lhardy@perimenterinstitute.ca}
\affiliation{%
 University of Waterloo, Departement of Physics and Astronomy \\
 Perimenter Institute for Theoretical Physics \\
 31 Caroline Street North, N2L 2Y5 \\
 Waterloo, Ontario
}%

\date{\today}

\begin{abstract}
A pair of interferometers can be coupled by allowing one path from each to overlap such that if the particles meet in this overlap region, they annihilate.  It was shown by one of us over thirty years ago that such annihilation-coupled interferometers can exhibit apparently paradoxical behaviour.  More recently, Bose \textit{et al}.\ and Marletto and Vedral have considered a pair of interferometers that are phase-coupled (where the coupling is through gravitational interaction).  In this case one path from each interferometer undergoes a phase-coupling interaction. We show that these phase-coupled interferometers exhibit the same apparent paradox as the annihilation-coupled interferometers, though in a curiously dual manner.    
\end{abstract}

\maketitle

\section{Introduction}

Over thirty years ago, Elitzur and Vaidman \cite{elitzur1993quantum} proposed a way to test for the presence of an object without (in a certain key sense) interacting with it. This proposal was dramatically illustrated by considering a bomb with a very sensitive trigger. The technique involved using a Mach-Zehnder (see Fig.\ \ref{fig:mach_zehnder}) interferometer tuned to have destructive interference in one output (the dark output).

The Elitzur-Vaidman paper inspired one of us to propose a gedanken experiment \cite{hardy1992quantum} with two overlapping interferometers, one for positrons, the other for electrons (see Fig.\ \ref{fig:hardy-original}), each tuned to have a dark output in the absence of the other. If the electron and positron meet on the overlapping paths, they annihilate. We will call these \emph{annihilation-coupled interferometers}. In this case, each interferometer measures the particle going through the other interferometer leading to an apparent paradox. This argument is, in fact, a proof of nonlocality without inequalities version of Bell's theorem \cite{bell1964einstein} (compare with the GHZ argument \cite{greenberger1990bell}). The apparent paradox is a consequence of implicit local reasoning.

More recently, Bose \textit{et al}.\ and Marletto-Vedral (B${}^+$MV)\cite{bose2016matter, bose2017spin,marletto2017grav} introduced an experiment also having two coupled interferometers but where, now, the coupling is a phase coupling due to gravitational interaction (see Fig.\ \ref{fig:bmv_like_setup}). We will call these \emph{phase-coupled interferometers}. This work has been the subject of much discussion and debate (\cite{roveli_qs_geometry_2019,anastopoulos2018comment-gme,belenchia2018quantum,cqg_hall_comment_2021,d2020classicalgme,cqg_reginatto_entanglement_2018,Danielson_2022,hall2018-comment-gme,Kent_2021,Marshman_2020,lami2023testing, martinmartinez2023gravity,no-go-theorem} and references therein).  

In the present paper, we will show how, with suitable tuning, the phase-coupled interferometers can exhibit an apparent paradox with exactly the same logical structure as in the annihilation-coupled interferometers case. This provides a rather striking illustration of the B${}^+$MV claim that the gravitational field must be non-classical.

We will see that the original annihilation-coupled interferometers and the phase-coupled interferometers implement this same apparent logical paradox in curiously dual manners from an interferometric point of view.

We can also implement this apparent logical paradox simply using two qubits prepared in a non-maximally entangled state as discussed in \cite{hardy1993nonlocality}.  The maximum fraction of cases where the apparent paradox is seen is $\frac{5\sqrt{5}-11}{2}$.  Neither of the coupled interferometers arrangements achieve this maximum.

\section{The Elitzur Vaidman bomb problem}

\begin{figure}[!ht]
\centering
\resizebox{0.4\textwidth}{!}{%
\begin{circuitikz}
\tikzstyle{every node}=[font=\LARGE]

\draw [, line width=1.6pt](-0.25,10.25) to[short] (9.75,10.25);
\draw [, line width=1.6pt](2,16.75) to[short] (2,10.25);
\draw [, line width=1.6pt](9.75,10.25) to[short] (9.75,18.25);
\draw [, line width=1.6pt](2,16.75) to[short] (11.25,16.75);

\node [font=\LARGE] at (-1.25,10.25) {$s$};
\node [font=\LARGE] at (5.75,11) {$v$};
\node [font=\LARGE] at (5.75,16.25) {$u$};
\node [font=\LARGE] at (10.5,16.25) {$c$};
\node [font=\LARGE] at (9,17.5) {$d$};

\draw [ line width=1.6pt, -Stealth] (0.75,10.25) -- (1.25,10.25);
\draw [ line width=1.6pt, -Stealth] (5.25,10.25) -- (5.75,10.25);
\draw [ line width=1.6pt, -Stealth] (5.25,16.75) -- (5.75,16.75);
\draw [ line width=1.6pt, -Stealth] (9.75,12.75) -- (9.75,13.25);
\draw [ line width=1.6pt, -Stealth] (2,13) -- (2,13.25);
\draw [ line width=1.6pt, -Stealth] (10,16.75) -- (10.5,16.75);
\draw [ line width=1.6pt, -Stealth] (9.75,17.25) -- (9.75,17.5);

\draw [line width=1.6pt, short] (10.25,10.75) -- (9.25,9.75);
\draw [line width=1.6pt, short] (2.5,17.25) -- (1.5,16.25);

\draw [line width=1.4pt, short] (2.5,17.25) -- (2,17.25);
\draw [line width=1.4pt, short] (2.25,17) -- (1.75,17);
\draw [line width=1.4pt, short] (2,16.75) -- (1.5,16.75);
\draw [line width=1.4pt, short] (1.75,16.5) -- (1.25,16.5);
\draw [line width=1.4pt, short] (1.5,16.25) -- (1,16.25);
\draw [line width=1.4pt, short] (10.25,10.75) -- (10.75,10.75);
\draw [line width=1.4pt, short] (10,10.5) -- (10.5,10.5);
\draw [line width=1.4pt, short] (9.75,10.25) -- (10.25,10.25);
\draw [line width=1.4pt, short] (9.5,10) -- (10,10);
\draw [line width=1.4pt, short] (9.25,9.75) -- (9.75,9.75);

\node [font=\LARGE] at (3,11.25) {$BS1$};
\node [font=\LARGE] at (8.5,15.55) {$BS2$};
\draw [line width=1.6pt, short] (2.5,10.75) -- (1.5,9.75);
\draw [line width=1.6pt, short] (10.25,17.25) -- (9.25,16.25);

\draw [, line width=1.6pt , dashed] (7.25,16.75) circle (1cm) node {\LARGE B} ;

\draw [, line width=1.6pt ] (11.25,17.75) rectangle  node {\LARGE $C$} (13.25,15.75);
\draw [, line width=1.6pt ] (8.75,20.25) rectangle  node {\LARGE $D$} (10.75,18.25);
\end{circuitikz}
}%
\caption{A Mach-Zehnder interferometer is tuned so that no particles are detected at $D$ if paths were unobstructed. $BS1$ has transmittance $t^2$ and reflectance $r^2$ whilst $BS2$ has these reversed.  A bomb whose trigger obstructs path $u$ can be placed at $B$ as shown.}
\label{fig:mach_zehnder}
\end{figure}
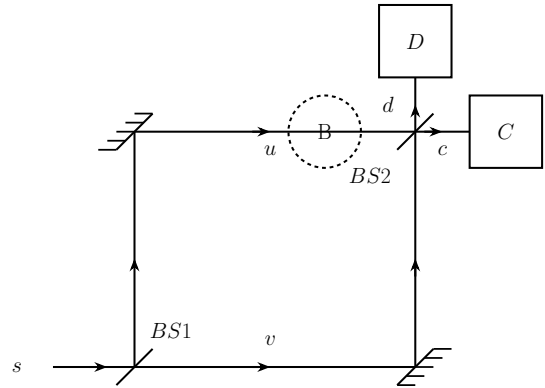 

A bomb-making factory produces bombs having a trigger so sensitive that, if a single particle of any type whatsoever impinges on it, the bomb will explode.  Sometimes, however, the trigger is missing.   The problem is to perform a test to see if the trigger is present.  In doing so it is OK if some bombs are exploded as long as, at the end of the day, we have some bombs which we know have a trigger but for which the bomb has not been exploded.

Elitzur and Vaidman's solution to this problem employs the Mach-Zehnder interferometer shown in Fig.\ \ref{fig:mach_zehnder}.  There is the possibility of placing an obstacle in path $u$ (this obstacle being taken to be the trigger of a bomb).  We assume the first beamsplitter has transmittance $t^2$ and reflectance $r^2$ (where $t$ and $r$ are real, $t^2+r^2=1$, and we take $t,r>0$) while the second beamsplitter has transmittance $r^2$ and reflectance $t^2$.   If we place a bomb with no trigger (so we have no obstacle) then the evolution is as follows
\[  |s\rangle \overset{BS1}{\longrightarrow}  ir|u\rangle + t|v\rangle  \overset{BS2}{\longrightarrow} ir( r|c\rangle +i t|d\rangle) + t(it|c\rangle + r|d\rangle) = i|c\rangle   \]
where we have included an $i$ factor when the particle is reflected off a beamsplitter (this guarantees that the transformation at each beamsplitter is unitary) and we have assumed that the total phase accumulated along each of the internal paths, $u$ and $v$, is a multiple of $2\pi$.  This means that detector $C$ will always fire and detector $D$ is \lq\lq dark".  Now we consider placing a bomb with its trigger protruding into path $u$ as shown so that it will absorb the particle and explode the bomb if the particle goes along path $u$
\[  |u\rangle |B\rangle \overset{P}{\longrightarrow} |\text{explosion}\rangle  \]
where $B$ stands for an unexploded bomb with a trigger and $P$ is the location of the trigger.  The evolution now goes as follows
\begin{align*}
|s\rangle |B\rangle &\overset{BS1,P}{\longrightarrow} ir|\text{explosion}\rangle + t|v\rangle |B\rangle \\ &\overset{BS2}{\longrightarrow} ir|\text{explosion}\rangle + t(it|c\rangle +r|d\rangle)|B\rangle
\end{align*}
Now we see that there is a probability $t^2r^2$ that detector $D$ will fire and the bomb will not explode (even though it has a trigger).  Since $D$ cannot fire if the trigger is missing then we have a way to deduce the presence of the trigger without exploding the bomb.  There is, of course, still a nonzero probability of exploding the bomb.  If detector $C$ fires then we do not know if the bomb has a trigger or not.  Elitzur and Vaidman showed how, by retesting cases where $C$ fires, it is possible to detect almost $50\% $ of bombs without exploding them in the limit. This limiting case requires taking $r$ to be very small but still non-zero (so, most times, the particle takes the path without the bomb).  An alternative arrangement, involving interrupted coherent evolution (the quantum Zeno effect) was proposed in Kwiat \textit{et al}.\ \cite{kwiat1995interaction} showing how it is possible to detect $(100-\delta)\% $ of the bombs for arbitrarily small $\delta$).

\section{Annihilation-coupled interferometers}

\begin{figure}[!ht]
\centering
\resizebox{0.45\textwidth}{!}{%
\begin{circuitikz}
\tikzstyle{every node}=[font=\huge]
\draw [, line width=1.5pt](19.25,8.25) to[short] (19.25,-1.5);
\draw[, line width=1.6pt] (21.25,6) to[short] (8.25,6);
\draw [, line width=1.6pt](8.25,6) to[short] (8.25,-1.5);
\draw[, line width=1.6pt] (19.25,-1.5) to[short] (6.25,-1.5);
\draw [line width=1.6pt, short] (9,6.75) -- (7.5,5.25);
\draw [line width=1.4pt, short] (9,6.75) -- (8.5,6.75);
\draw [line width=1.4pt, short] (8.75,6.5) -- (8.25,6.5);
\draw [line width=1.4pt, short] (8.5,6.25) -- (8,6.25);
\draw [line width=1.4pt, short] (8.25,6) -- (7.75,6);
\draw [line width=1.4pt, short] (8,5.75) -- (7.5,5.75);
\draw [line width=1.4pt, short] (7.75,5.5) -- (7.25,5.5);
\draw [line width=1.4pt, short] (7.5,5.25) -- (7,5.25);
\draw [line width=1.6pt, short] (20,-0.75) -- (18.5,-2.25);
\draw [line width=1.4pt, short] (20,-0.75) -- (20.5,-0.75);
\draw [line width=1.4pt, short] (19.75,-1) -- (20.25,-1);
\draw [line width=1.4pt, short] (19.5,-1.25) -- (20,-1.25);
\draw [line width=1.4pt, short] (19.25,-1.5) -- (19.75,-1.5);
\draw [line width=1.4pt, short] (19,-1.75) -- (19.5,-1.75);
\draw [line width=1.4pt, short] (18.75,-2) -- (19.25,-2);
\draw [line width=1.4pt, short] (18.5,-2.25) -- (19,-2.25);
\draw [line width=1.6pt, short] (9,-0.75) -- (7.25,-2.25);
\draw [line width=1.6pt, short] (20.25,6.75) -- (18.25,5);

\draw[, line width=1.6pt] (11.75,2.75) to[short] (3.5,2.75);
\draw [, line width=1.6pt](3.5,13.5) to[short] (3.5,1);
\draw [, line width=1.6pt](3.5,13.5) to[short] (13.5,13.5);
\draw [, line width=1.6pt](11.75,2.75) to[short] (11.75,15.5);

\draw [line width=1.6pt, short] (12.5,3.5) -- (11 ,2);
\draw [line width=1.4pt, short] (12.5,3.5) -- (13,3.5);
\draw [line width=1.4pt, short] (12.25,3.25) -- (12.75,3.25);
\draw [line width=1.4pt, short] (12 ,3) -- (12.5 ,3);
\draw [line width=1.4pt, short] (11.75,2.75) -- (12.25,2.75);
\draw [line width=1.4pt, short] (11.5 ,2.5) -- (12 ,2.5);
\draw [line width=1.4pt, short] (11.25 ,2.25) -- (11.75 ,2.25);
\draw [line width=1.4pt, short] (11  ,2 ) -- (11.5 ,2);

\draw [line width=1.6pt, short] (4.5,14.5) -- (3,13);
\draw [line width=1.4pt, short] (4.5,14.5) -- (4,14.5);
\draw [line width=1.4pt, short] (4.25,14.25) -- (3.75,14.25);
\draw [line width=1.4pt, short] (4 ,14 ) -- (3.5,14 );
\draw [line width=1.4pt, short] (3.75 ,13.75 ) -- (3.25,13.75 );
\draw [line width=1.4pt, short] (3.5 ,13.5 ) -- (3 ,13.5 );
\draw [line width=1.4pt, short] (3.25 ,13.25 ) -- (2.75,13.25 );
\draw [line width=1.4pt, short] (3 ,13 ) -- (2.5,13 );
\draw [line width=1.4pt, short] (4.25,3.5) -- (2.5,2);
\draw [line width=1.4pt, short] (12.5,14.25) -- (10.75,12.75);

\draw [ line width=1.6pt, -Stealth] (6.75,-1.5) -- (7,-1.5);
\draw [ line width=1.6pt, -Stealth] (13.25,-1.5) -- (13.5,-1.5);
\draw [ line width=1.6pt, -Stealth] (19.25,1.25) -- (19.25,1.75);
\draw [ line width=1.6pt, -Stealth] (19.25,7) -- (19.25,7.25);
\draw [ line width=1.6pt, -Stealth] (19.75,6) -- (20.5,6);
\draw [ line width=1.6pt, -Stealth] (13.5,6) -- (13.75,6);
\draw [ line width=1.6pt, -Stealth] (8.25,1.25) -- (8.25,2);
\draw [ line width=1.6pt, -Stealth] (3.5,1.75) -- (3.5,2.25);
\draw [ line width=1.6pt, -Stealth] (3.5,7.75) -- (3.5,8);
\draw [ line width=1.6pt, -Stealth] (6.5,2.75) -- (7.25,2.75);
\draw [ line width=1.6pt, -Stealth] (11.75,7.75) -- (11.75,8);
\draw [ line width=1.6pt, -Stealth] (6.75,13.5) -- (7.75,13.5);
\draw [ line width=1.6pt, -Stealth] (11.75,14.25) -- (11.75,14.5);
\draw [ line width=1.6pt, -Stealth] (12.25,13.5) -- (12.75,13.5);
\draw [ fill={rgb,255:red,0; green,0; blue,0} , line width=1.6pt ] (8.25,2.75) circle (0.25cm);
\node [font=\huge] at (8.75,3.5) {P};
\node [font=\huge] at (10.5,8) {$u^+$};
\node [font=\huge] at (4.5,8) {$v^+$};
\node [font=\huge] at (12.5,12.5) {$d^+$};
\node [font=\huge] at (11,14.5) {$c^+$};
\node [font=\huge] at (13.75,-0.75) {$v^-$};
\node [font=\huge] at (13.75,5) {$u^-$};
\node [font=\huge] at (20.5,5.5) {$c^-$};
\node [font=\huge] at (18.25,6.75) {$d^-$};
\draw [, line width=1.6pt ] (11,17.5) rectangle  node {\huge $C^+$} (13,15.5);
\draw [, line width=1.6pt ] (13.5,14.5) rectangle  node {\huge $D^+$} (15.5,12.5);
\draw [, line width=1.6pt ] (18.25,10.25) rectangle  node {\huge $D^-$} (20.25,8.25);
\draw [, line width=1.6pt ] (21.5,7) rectangle  node {\huge $C^-$} (23.5,5);
\node [font=\huge] at (4.3,1.4) {$s^+$};
\node [font=\huge] at (5.5,-1.5) {$s^-$};
\draw [ , dashed] (15,7) rectangle node {\LARGE $U^-$} (17,5);
\draw [ , dashed] (11,12) rectangle node {\LARGE $U^+$} (13,10);
\end{circuitikz}
}%

\caption{Annihilation coupled interferometers.  One electron and one positron go through two interferometers. If they meet at the point $P$, they will annihilate.   The experimenter can decide whether to insert detectors $U^+$ and $U^-$. The first beamsplitter, $BS1^\pm$ in each interferometer, has transmittance $t^2$ and reflectance $r^2$ whilst the second beamsplitters, $BS2^\pm$, have these reversed.}
\label{fig:hardy-original}
\end{figure}
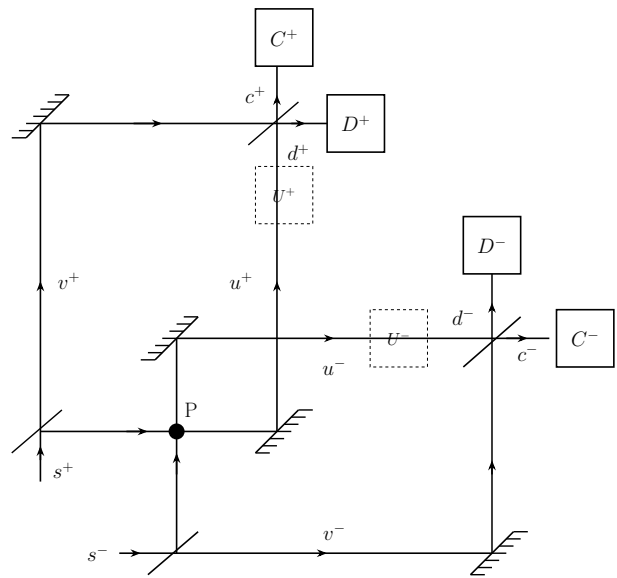

The annihilation-coupled interferometers are shown in Fig. \ref{fig:hardy-original}.   Here we have one interferometer for positrons and one for electrons. We tune the interferometers as we did for the Elitzur-Vaidman case so that no particles arrive at the $D^\pm$ detector in the case that the interferometers are not coupled.  Now we introduce coupling such that
\[  |u^+\rangle|u^-\rangle \overset{P}{\longrightarrow} |\gamma\rangle  ~~\text{at}~~P \]
where $|\gamma\rangle$ is the radiation resulting from annihilation.  Taking this into account, the evolution goes as follows
\begin{align}
|s^+\rangle|s^-\rangle \overset{\text{BS1}^\pm}{\longrightarrow}& (ir|u^+\rangle+t|v^+\rangle) (ir|u^-\rangle+t|v^-\rangle) \nonumber\\
\overset{P}{\longrightarrow}& ir^2|\gamma\rangle + irt|u^+\rangle|v^-\rangle  \nonumber \\
&~ + irt|v^+\rangle |u^-\rangle
+ t^2 |v^+\rangle |v^-\rangle   \label{afterP}
\end{align}
We have the option to place, or not place, a detector $U^\pm$ in path $u^\pm$ as shown in Fig. \ref{fig:hardy-original}. We will put $U^+=1$ when detector $U^+$ fires and similarly for the other detectors.  We consider four different experiments (out of order for later comparison).
\begin{description}
  \item[Four. $U^+$ and $U^-$ in place] We see that
  \begin{equation} U^+=1 ~~\text{and}~~ U^-=1~~\text{never happens} \end{equation}
  because there is no $|u^+\rangle|u^-\rangle$ term in \eqref{afterP}
  \item[Two. $D^+$ in place, $U^-$ absent] We see that
  \begin{equation} D^+=1 ~~~\Rightarrow ~~  U^-=1  \end{equation}
  because, evolving \eqref{afterP} through $BS2^+$ gives
  \begin{align*}
  & ir^2|\gamma\rangle +irt(r|c^+\rangle + it|d^+\rangle) |v^-\rangle \\
  & + (it|c^+\rangle + r|d^+\rangle)(irt|u^-\rangle + t^2|v^-\rangle)
  \end{align*}
  for which the coefficients in front of the $|d^+\rangle|v^-\rangle$ term cancel (this is exactly the Elitzur-Vaidman trick).
  \item[Three. $U^+$ absent, $U^-$ in place] We see that
  \begin{equation}  U^+=1 ~~ \Leftarrow ~~ D^-=1   \end{equation}
  by evolving \eqref{afterP} through BS2${}^-$ (this is clearly true by symmetry comparing with the previous case).
  \item[One. $U^+$ and $U^-$ absent] We see that
  \begin{equation}  D^+=1 ~~\text{and}~~ D^-=1 ~~\text{happens sometimes} \end{equation}
  because, when we evolve \eqref{afterP} through both BS2${}^\pm$ the $|d^+\rangle|d^-\rangle$ term has coefficient $-t^2r^2$ -so the probability for this event is $r^4t^4$.   The maximum value for this probability is $\frac{1}{16}$ and happens when $r=\frac{1}{\sqrt{2}}$.
\end{description}
These results appear paradoxical. Consider a case where we do experiment one and obtain $D^+=1$ and $D^-=1$.  This implies, from experiment two, that we would have got $U^-=1$, and, from experiment three, that we would have got $U^+=1$.  This appears to contradict the prediction for experiment four which tells us we cannot get $U^+=1$ and $U^-=1$.

\section{Phase-coupled interferometers}\label{sec:phasecoupledinterferometers}

\begin{figure}[!ht]
\centering
\resizebox{0.4\textwidth}{!}{%
\begin{circuitikz}
\tikzstyle{every node}=[font=\LARGE]

\draw [ color={rgb,255:red,200; green,50; blue,40} , fill={rgb,255:red,155; green,193; blue,253}, line width=1.6pt ] (2.1,12.1) rectangle node {\LARGE Phase Interaction} (9.6,7.9);

\draw [, line width=1.6pt](-0.25,11.25) to[short] (9.75,11.25);
\draw [, line width=1.6pt](2,17.75) to[short] (2,11.25);
\draw [, line width=1.6pt](2,17.75) to[short] (11.5,17.75);
\draw [, line width=1.6pt](9.75,11.25) to[short] (9.75,19.25);

\draw [, line width=1.6pt](-0.25,9) to[short] (9.75,9);
\draw [, line width=1.6pt](2,9) to[short] (2,2.5);
\draw [, line width=1.6pt](2,2.5) to[short] (11.5,2.5);
\draw [, line width=1.6pt](9.75,9) to[short] (9.75,1.25);

\draw [line width=1.6pt, short] (10.25,11.75) -- (9.25 ,10.75);
\draw [line width=1.4pt, short] (10.25,11.75) -- (10.75,11.75);
\draw [line width=1.4pt, short] (10,11.5) -- (10.5,11.5);
\draw [line width=1.4pt, short] (9.75,11.25) -- (10.25,11.25);
\draw [line width=1.4pt, short] (9.5,11) -- (10,11);
\draw [line width=1.4pt, short] (9.25,10.75) -- (9.75,10.75);
\draw [line width=1.6pt, short] (2.5,18.25) -- (1.5,17.25);
\draw [line width=1.4pt, short] (2.5,18.25) -- (2,18.25);
\draw [line width=1.4pt, short] (2.25,18) -- (1.75,18);
\draw [line width=1.4pt, short] (2,17.75) -- (1.5,17.75);
\draw [line width=1.4pt, short] (1.75,17.5) -- (1.25,17.5);
\draw [line width=1.4pt, short] (1.5,17.25) -- (1,17.25);

\draw [line width=1.6pt, short] (10.25,18.25) -- (9.25,17.25);
\draw [line width=1.6pt, short] (2.5,11.75) -- (1.5,10.75);

\draw [line width=1.6pt, short] (2.5,2) -- (1.5,3);
\draw [line width=1.4pt, short] (2.5,2) -- (2,2);
\draw [line width=1.4pt, short] (2.25,2.25) -- (1.75,2.25);
\draw [line width=1.4pt, short] (2,2.5) -- (1.5,2.5);
\draw [line width=1.4pt, short] (1.75,2.75) -- (1.25,2.75);
\draw [line width=1.4pt, short] (1.5,3) -- (1,3);
\draw [line width=1.6pt, short] (10.25,8.5) -- (9.25,9.5);
\draw [line width=1.4pt, short] (10.75,8.5) -- (10.25,8.5);
\draw [line width=1.4pt, short] (9.75,9.5) -- (9.25,9.5);
\draw [line width=1.4pt, short] (10,9.25) -- (9.5,9.25);
\draw [line width=1.4pt, short] (10.25,9) -- (9.75,9);
\draw [line width=1.4pt, short] (10.5,8.75) -- (10,8.75);

\draw [line width=1.6pt, short] (2.5,8.5) -- (1.5,9.5);
\draw [line width=1.6pt, short] (10.25,2) -- (9.25,3);

\draw [, line width=1.6pt ] (8.75,21.25) rectangle  node {\LARGE $D_1$} (10.75,19.25);
\draw [, line width=1.6pt ] (11.5,18.75) rectangle  node {\LARGE $C_1$} (13.5,16.75);
\draw [, line width=1.6pt ] (8.75,1.25) rectangle  node {\LARGE $D_2$} (10.75,-0.75);
\draw [, line width=1.6pt ] (11.5,3.5) rectangle  node {\LARGE $C_2$} (13.5,1.5);
\draw [ , dashed] (6.25 ,18.75) rectangle node {\LARGE $U_1$} (8.25,16.75);
\draw [ , dashed] (6.25,3.5) rectangle node {\LARGE $U_2$} (8.25,1.5);

\node [font=\LARGE] at (-0.75,11.25) {$s_1$};
\node [font=\LARGE] at (-0.75,9) {$s_2$};
\node [font=\LARGE] at (5,11.75) {$v_1$};
\node [font=\LARGE] at (5,17.25) {$u_1$};
\node [font=\LARGE] at (5,8.25) {$v_2$};
\node [font=\LARGE] at (5,3.25) {$u_2$};
\node [font=\LARGE] at (10.5,17.25) {$c_1$};
\node [font=\LARGE] at (9,18.5) {$d_2$};
\node [font=\LARGE] at (10.5,3) {$c_2$};
\node [font=\LARGE] at (9,1.75) {$d_2$};
\draw [ line width=1.6pt, -Stealth] (0.75,9) -- (1.25,9);
\draw [ line width=1.6pt, -Stealth] (5.25,9) -- (5.75,9);
\draw [ line width=1.6pt, -Stealth] (9.75,6.25) -- (9.75,5.25);
\draw [ line width=1.6pt, -Stealth] (2,6) -- (2,5.25);
\draw [ line width=1.6pt, -Stealth] (5.5,2.5) -- (6,2.5);
\draw [ line width=1.6pt, -Stealth] (10.5,2.5) -- (10.75,2.5);
\draw [ line width=1.6pt, -Stealth] (9.75,2) -- (9.75,1.75);
\draw [ line width=1.6pt, -Stealth] (0.75,11.25) -- (1.25,11.25);
\draw [ line width=1.6pt, -Stealth] (5.25,11.25) -- (5.75,11.25);
\draw [ line width=1.6pt, -Stealth] (5.25,17.75) -- (5.75,17.75);
\draw [ line width=1.6pt, -Stealth] (9.75,13.75) -- (9.75,14.25);
\draw [ line width=1.6pt, -Stealth] (2,14) -- (2,14.25);
\draw [ line width=1.6pt, -Stealth] (10.25,17.75) -- (10.75,17.75);
\draw [ line width=1.6pt, -Stealth] (9.75,18.5) -- (9.75,18.75);


\end{circuitikz}
}%

\caption{Phase coupled interferometers setup.  If the particles take paths $v_1$ and $v_2$, then there is a phase coupling as described by \eqref{phasecoupling}.  The first beamsplitter in each interferometer ($BS1^{1,2}$) has transmittance $t^2$ and reflectance $r^2$ whilst the second beamsplitter ($BS2^{1,2}$) has these reversed.  Experimenters can decide whether to do insert detectors $U_1$ and $U_2$ or not.}
\label{fig:bmv_like_setup}
\end{figure}
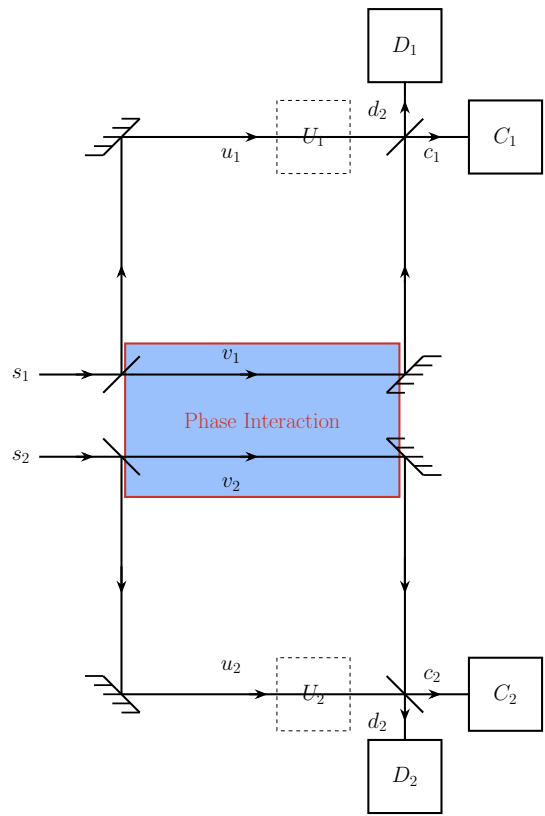

Now we consider the phase-coupled interferometers of B${}^+$MV shown in Fig.\ \ref{fig:bmv_like_setup} where the interferometers are each tuned as in the Elitzur-Vaidman case so that, when there is no coupling, detectors $D_1$ and $D_2$ never click .  The phase-coupling interaction is as follows
\begin{equation}\label{phasecoupling}
|v_1\rangle|v_2\rangle \longrightarrow e^{i\phi} |v_1\rangle|v_2\rangle
\end{equation}
Note that, this time, we couple on the $v^{1,2}$ paths.  This interaction can be achieved by gravitational interaction as described by B${}^+$MV or it could be achieved by sending the particles through an appropriate nonlinear phase coupling medium.  The evolution goes as follows
\begin{align}
|s_1\rangle|s_2\rangle
\overset{\text{BS}1^{1,2}}{\longrightarrow}& (ir|u_1\rangle+t|v_1\rangle) (ir|u_2\rangle+t|v_2\rangle) \nonumber\\
\overset{\phi}{\longrightarrow} &
-r^2|u_1\rangle|u_2\rangle + irt|u_1\rangle|v_2\rangle \nonumber \\
& + irt|v_1\rangle |u_2\rangle
 + t^2 e^{i\phi}|v_1\rangle |v_2\rangle  \label{afterphi}
\end{align}
We have the option to put $U_{1,2}$ in place or not.  We consider four experiments.
\begin{description}
  \item[One. $U_1$ and $U_2$ in place]  We see that
  \begin{equation}
  U_1=1 ~~\text{and}~~ U_2=1 ~~\text{happens sometimes}
  \end{equation}
  The $|u_1\rangle|u_2\rangle$ term in \eqref{afterphi} has coefficient $-r^2$ and hence the probability for this is $r^4$.
  \item[Two. $U_1$ in place and $U_2$ absent] We see that
  \begin{equation}
  U_1=1~~ \Rightarrow ~~ C_2=1
  \end{equation}
  This is because, if $U_1$ fires, then the interferometers are decoupled and so $D_2$ cannot fire.  Mathematically we can see this by evolving \eqref{afterphi} through BS$2_2$ giving
  \begin{align*}
  & (-r^2|u_1\rangle+ irt|v_1\rangle) (r|c_2\rangle +it|d_2\rangle) \\
  & +(irt|u_1\rangle+ t^2e^{i\phi}|v_1\rangle)(it|c_2\rangle +r|d\rangle)
  \end{align*}
  We see that the coefficients in front of the $|u_1\rangle|d_2\rangle$ cancel.
  \item[Three. $U_1$ absent and $U_2$ in place] We see that
  \begin{equation}
  C_1 ~~ \Leftarrow ~~ U_2
  \end{equation}
  because, if $U_2$ fires, then the interferometers are decoupled and so $D_1$ cannot fire.
  \item[Four. $U_1$ and $U_2$ absent] Then we can tune the parameters, $r$ and $t$ such that
  \begin{equation}
  C_1 =1  ~~\text{and}~~ C_2=1~~\text{never happens}
  \end{equation}
  To see this we note that, after evolving through BS$2_1$ and $BS2_2$, the coefficient in front of the $|c_1\rangle|c_2\rangle$ term is
  \begin{equation}
  -(r^4 + 2r^2t_2 + t^4 e^{i\phi})
  \end{equation}
  which we can set to zero by, for example, putting $\phi=\pi$ and $r^2=\frac{2-\sqrt{2}}{2}$ (with $t^2=1-r^2$).
\end{description}
These predictions have exactly the same logical structure as experiments one to four with the annihilation-coupled interferometers.  Consider a case where we do experiment one and obtain $U_1=1$ and $U_2=1$. Then experiments two and thee imply, respectively, that we would have obtained $C_2=1$ and $C_1=1$ but this appears to contradict experiment four.  

With the choices for $r^2,t^2$ and $\phi$ just given, the probability for both detectors $U_1$ and $U_2$ firing (equal to $r^4$) is
\begin{equation}
p_\text{max}(U_1U_2)= \frac{(3-\sqrt{2})}{2} \approx 0.0857
\end{equation}
This is the maximum value for this probability if we tune so that $p(C_1C_2)=0$.
Interestingly, this is less than the maximum probability for this apparent logical paradox for two qubits prepared in an arbitrary entangled state as studied in \cite{hardy1993nonlocality} wherein the maximum probability was $\frac{5\sqrt{5}-11}{2}\approx 0.0902$ (Mermin \cite{mermin1998ithaca} pointed out that this is equal to $\tau^{-5}$ where $\tau$ is the golden mean).

Predictions for experiments one to four here have the same logical structure as those for experiments one to four in the annihilation-coupled case.  However, the way in which these results are obtained is, in some sense, dual to that case.

\section{Nonlocality}\label{sec:nonlocality}

The apparent paradox discussed here can be understood as a proof of Bell's theorem, that nonlocal hidden variable theories cannot reproduce the predictions of Quantum Theory. Consider the phase-coupled interferometer example.  Assume that hidden variables, $\lambda$, are shared when the particles interact and that these hidden variables locally determine probabilities for the different detector events.  Assume, further, that probabilities factorize when conditioned on $\lambda$.  There must be $\lambda\in S$ such that $p(U_1=1, U_2=1|\lambda)>0$ for all $\lambda\in S$. For $\lambda\in S$, we must have $p(C_2=1|\lambda)=1$ since we need to be sure to agree with the predictions for experiment two.  Similarly, we must have $p(C_1=1|\lambda)=1$ for $\lambda\in S$ to be sure to agree with the predictions for experiment one.  Hence, for $\lambda\in S$, we must have $p(C_1=1,C_2=1|\lambda)=p(C_1=1|\lambda)p(C_2=1|\lambda)=1$.  However this contradicts experiment four which dictates that $p(C_1=1,C_2=1|\lambda)=0$ for all $\lambda$.

\begin{figure}[!ht]
    \centering
    \includegraphics[width =0.46\textwidth ]{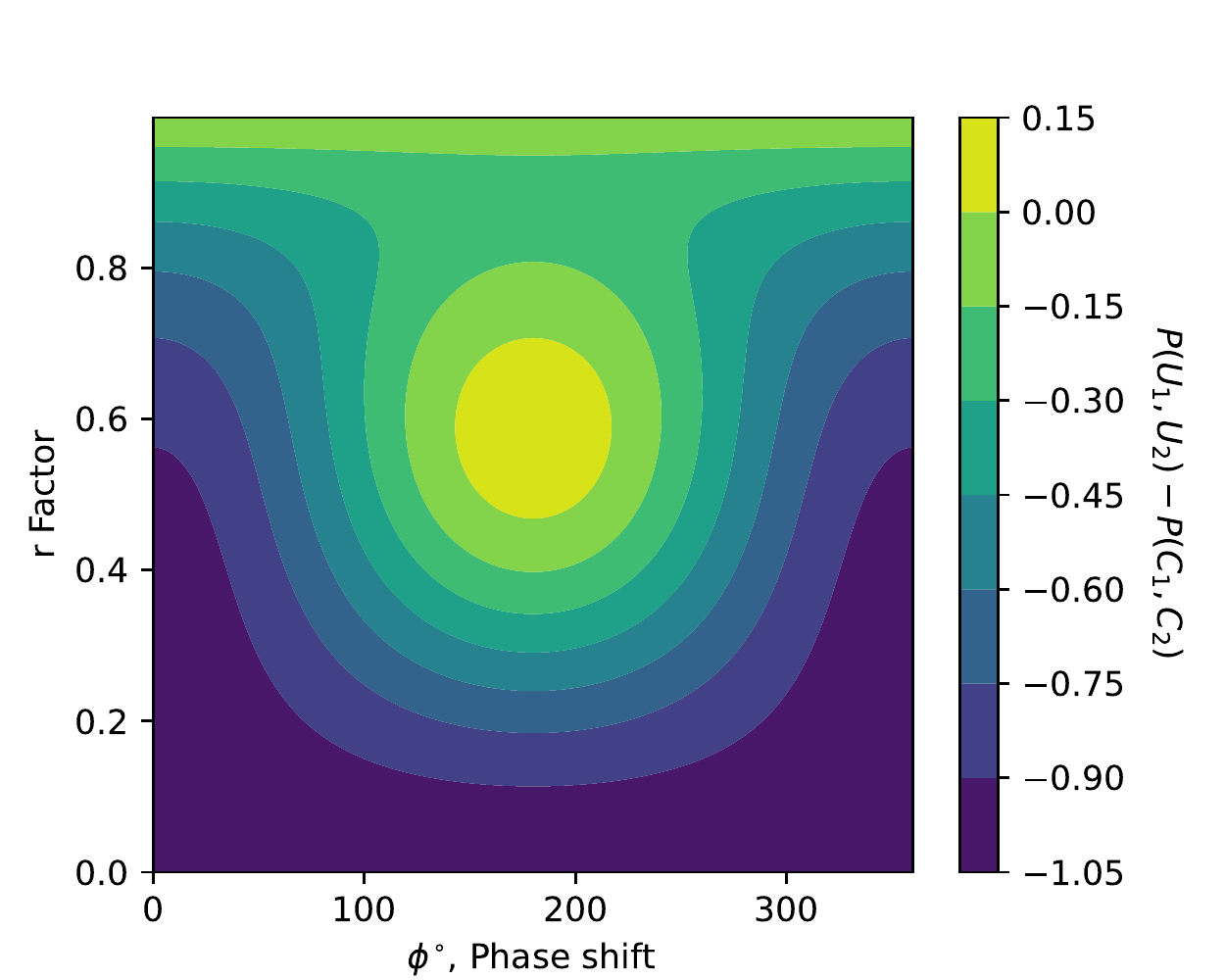}
    \caption{Plot shows the factor $r$ versus the gravitational phase shift factor $\phi$. Colors indicates $P(U_1,U_2) - P(C_1,C_2)$. For $P(U_1,U_2) - P(C_1,C_2) > 0$ (yellow middle circle) we see a direct contradiction with the local hidden variable models in the described experiments.}
    \label{fig:full_solution}
\end{figure}

With the phase-coupled interferometers, it is natural to consider the case where, in experiment four, we allow $p(C_1=1,C_2=1)$ to be greater than zero.  As long as $p(U_1=1,U_2=1)>p(C_1=1,C_2=1)$, we still have an apparent paradox since then experiment one, two, and three imply more cases where $C_1=1$ and $C_2=1$ than are seen in experiment four.  In fact, it is easy to derive \cite{mermin1994quantum, garuccio1995hardy} the following Bell inequality 
\begin{equation} \label{bellinequality}
p(U_1,U_2) - p(U_1,\overline{C}_2) - p(\overline{C}_1,U_2) - p(C_1,C_2) \leq 0  
\end{equation}
(a \lq\lq logical" derivation is given in Appendix \ref{app:bellinequality} using techniques from \cite{hardy1991new,abramsky2012logical}) where $C_1$ indicates detector $C_1$ fires and $\overline{C}_1$ indicates it does not fire. For the phase-coupled interferometers the middle two probabilities are always equal to zero so the violation of this inequality is equal to $p(U_1,U_2)-p(C_1,C_2)$. We plot this violation as we vary $r$ and $\phi$ in Fig.\ref{fig:full_solution}. The maximum violation is $0.0990$ when $r=0.58309$ and $\phi=\pi$.

\section{Experiments}

The correlations in this apparent paradox have been implemented in many experiments (here is an incomplete list \cite{torgerson1995experimental,boschi1997ladder,hessmo2004experimental,irvine2005realization,carlson2006quantum,lundeen2009experimental,yokota2009direct,guo2015experimentally,das2017experimental,luo2018experimental,yang2019stronger}).   Three of these experiments were able to mirror the interferometric structure of annihilation coupling.  Irvine \textit{et al}. \cite{irvine2005realization} and Yokota \textit{et al}.\ \cite{yokota2009direct} used the Hong Ou Mandel effect \cite{hong1987measurement} at a beamsplitter to implement effective annihilation (as proposed in \cite{hardy1992aquantum}).  Lundeen and Steinberg \cite{lundeen2009experimental} use an interference effect that acts as an absorptive two photon switch to implement annihilation.  Two of these interferometric experiments \cite{lundeen2009experimental, yokota2009direct} also implemented weak measurements to test the ideas of Aharanov \textit{et al}.\ \cite{aharonov2002revisiting} (see also the work by Vaidman \cite{vaidman1993lorentz}).    

It would be interesting to implement the phase-coupled interferometers experiment.  Of most interest would be to implement the gravitational version of this experiment since this has implications for Quantum Gravity.  However, an implementation using more experimentally feasible phase coupling technique would be of interest in the short term.  Resch, Ludeen, and Steinberg \cite{resch2002conditional} implemented such a phase coupling in a quantum optical setting which might be suitable to implement phase-coupled interferometers.  Further, it would be interesting to provide a weak measurement analysis of the phase-coupled case to contrast with the above work in the annihilation-coupled case.

\section{On the Quantum nature of gravity}

As shown by B${}^+$MV, the phase coupling can be due to gravitational interaction.  In this case, we have
\[ \phi = \frac{Gm^2L}{\hbar d} \]
where $L$ is the length of the arms that interact, and $d$ is the distance between them.  

Nonlocal correlations, such as those discussed in Sec.\ \ref{sec:phasecoupledinterferometers}, can be regarded as a witness for quantum entanglement.  It is not possible to generate entanglement between two parties with local operations and classical communication (see reviews \cite{horodecki2009quantum,chitambar2014everything} and references therein).  This suggests (as argued by B${}^+$MV) that the gravitational interaction between the two interferometers must be quantum. 

We could try to go further.  It has been argued by Kent \cite{Kent2009} that General Relativity cannot produce nonlocal correlations since it is a local field theory (though see Sec.\ 31.2 of \cite{hardy2016operational} for a tentative counterview stemming from the fact that operationally local observables cannot be thought of as being localized on the manifold).  However, here we see nonlocal correlations, so we can argue that these correlations cannot have been produced by a classical general relativistic interaction.   For this argument to work, we need to spell out what is being assumed.  Let two systems, $A$ and $B$ (which may be quantum), not interact with each other directly, but each is allowed to interact with a third system, $G$, which we take to be the gravitational field.  System $A$ has a setting $S_A$ and an outcome $O_A$ and system $B$ has a setting $S_B$ and an outcome $O_B$.  System $G$ may have some variable property, $\phi_G$, that we can set.   Then, we are interested in a notion of classicality for system $G$ for which the joint probability for $O_AO_B$ is given by
\begin{align}
\label{ABGmodel}
p(O_AO_B|S_AS_B\phi_G)& = \nonumber \\
\int \rho(\lambda)d\lambda p(O_A|& S_A\phi_G,\lambda) p(O_B|S_B\phi_G,\lambda) 
\end{align}
where $\lambda$ are some extra variables we may need to fully model the interaction, and $\rho(\lambda)$ is a probability distribution over $\lambda$.  
In this case, the interaction between $A$ and $B$ happens by sharing the variables $\phi_G$ and $\lambda$.  We can use \eqref{ABGmodel} to derive Bell inequalities (such as \eqref{bellinequality}).  Hence, models for which \eqref{ABGmodel} is true cannot account for nonlocal correlations such as those discussed in this paper.   

Galley \textit{et al}.\ \cite{no-go-theorem} show, in the context of general probabilistic theories, that the quantum predictions for such experiments are inconsistent with $G$ being classical.  The hybrid quantum-classical model of Oppenheim and collaborators \cite{oppenheim2018post,oppenheim2022gravitationally}, where gravity is treated classically, is an example of a model that would be ruled out if the quantum predictions for a B${}^+$MV experiment were seen.   

\section{Acknowledgments}

This work is supported by Perimeter Institute for Theoretical Physics. Research at Perimeter Institute is supported by the Government of Canada through Industry Canada and by the Province of Ontario through the Ministry of Research and Innovation.  LH acknowledges support of the ID\# 61466 grant from the John Templeton Foundation, as part of the \lq\lq Quantum Information Structure of Spacetime (QISS)'' project (qiss.fr). Most of the graphics in this work have been generated using TikZ. Perimeter Institute acknowledges that it is situated on the traditional territory of the Anishinaabe, Haudenosaunee, and Neutral peoples.  We thank the Anishinaabe, Haudenosaunee, and Neutral peoples for hosting us on their land.

\bibliography{bib}
\bibliographystyle{plain}

\appendix

\section{Bell inequality}\label{app:bellinequality}

 Here we provide a derivation of the Bell inequality in \eqref{bellinequality} using the logical approach of \cite{hardy1991new, abramsky2012logical}.  Let $\{ S_n : n=1 ~\text{to}~N\}$ be a set of logical statements.  Imagine we are running an experiment many times in which each of these statements may be true.   Let $p(S_n)$ be the probability that statement $S_n$ is true.   Then the probability, $P_\text{all}$, are true is bounded by
\[ 1-P_\text{all} \leq \sum_{n=1}^N 1-p(S_n)  \]
using De Morgan's law 
\[ \lnot\left(\underset{n}{\bigwedge} S_n\right)=\underset{n}{\bigvee}(\lnot S_n) \]
If the set of statements are incompatible in some model then $P_\text{all}=0$.  This gives us the inequality
\begin{equation} \label{logicalinequality} 
\sum_{n=1}^N p(S_n) \leq N-1  
\end{equation}
If the set of statements are incompatible in local hidden variable theories then this is a Bell inequality.  
The following four logical statements are incompatible in a local hidden variable theory (see Sec.\ \ref{sec:nonlocality})
\begin{align*}
 S_1 &~~~~~~~ U_1 \land U_2 \\
 S_2 &~~~~~~~ \lnot ( U_1 \land \overline{C}_2) \\
 S_3 &~~~~~~~ \lnot (\overline{C}_1 \land U_2) \\
 S_4 &~~~~~~~ \lnot (C_1 \land C_2)   
\end{align*}
Using $p(\lnot( U_1 \land \overline{C}_2))=1-p(U_1\overline{C}_2)$ (and similar expressions) in \eqref{logicalinequality} gives us \eqref{bellinequality}.

\end{document}